\documentclass[twocolumn]{emulateapj}
\slugcomment{Submitted to ApJ}
\shorttitle{QHC21}
\shortauthors{T. Kojo et al.}
\newcommand{\Slash}[1]{{\ooalign{\hfil$#1$\hfil\crcr\raise.167ex\hbox{/}}}}

\newcommand \beq{\begin{eqnarray}}
\newcommand \eeq{\end{eqnarray}}

\newcommand{\blueflag}[1]{{\color{blue} #1}}

\usepackage[colorlinks=true,linktocpage=true,linkcolor=blue,citecolor=blue]{hyperref}

\usepackage{ulem}

\begin{document}
\title{ Implications of NICER for neutron star matter: the QHC21 equation of state }
\author{Toru Kojo}
\affiliation{Key Laboratory of Quark and Lepton Physics (MOE) and Institute of Particle Physics, Central China Normal University, Wuhan 430079, China} 
\affiliation{Department of Physics, Tohoku University, Sendai 980-8578, Japan}  
\author{Gordon Baym}
\affiliation{Illinois Center for Advanced Studies of the Universe}
\affiliation{Department of Physics, University of Illinois at Urbana-Champaign, 1110 W. Green Street, Urbana, IL 61801, USA}
\affiliation{RIKEN iTHEMS Program, Wako, Saitama 351-0198, Japan}  
\author{Tetsuo Hatsuda}
\affiliation{RIKEN iTHEMS Program, Wako, Saitama 351-0198, Japan\\}

\begin{abstract}

  The recent NICER measurement of the radius of the neutron star PSR J0740+6620, and the inferred small variation of radii from 1.4$M_\odot$ to 2.1$M_\odot$, reveal key features of the equation of state of neutron star matter.  The pressure rises rapidly in the regime of baryon density  $n \sim$ 2-4 times nuclear saturation density, $n_0$ --  the region where we expect hadronic matter to be undergoing transformation into quark matter --  and the pressure in the nuclear regime is greater than predicted by microscopic many-body variational calculations of nuclear matter.   To incorporate these insights into  the microscopic  physics from the nuclear to the quark matter regimes, 
 we construct an equation of state, QHC21, within the framework of quark-hadron crossover (QHC).
 We include nuclear matter results primarily based on the state-of-the-art chiral effective field theory, but also note results of using nuclear matter variational calculations based on empirical nuclear forces.  We employ explicit nuclear degrees of freedom only up to $n \sim 1.5n_0$,        
in order to explore the possibility of 
further physical degrees of freedom than nucleonic here. 
The resulting QHC21, which has  a peak in sound velocity in $\sim 2$-$4 n_0$,
is stiffer than the earlier QHC19 below 2$n_0$,  predicting larger radii in substantial agreement with the NICER data.
\\

\end{abstract}

\section{Introduction}
\label{sec:introduction}

 Recent data from the NICER observatory has yielded the important inference that the equatorial radius of the most massive neutron star known,  PSR J0740+6620, with mass $M/M_\odot = 2.08 \pm 0.07$ \citep{Fonseca:2021wxt}, is 13.7$^{+2.6}_{-1.5}$ km,  as analyzed by \cite{Miller:2021qha} combining NICER and XMM data for PSR J0740+6620. With nuclear physics constraints at low density, and gravitational radiation data from GW170817 added in, Miller et al.'s inferred radius drops to  $12.35 \pm 0.75$ km for a 2.08 $M_\odot$ neutron star.   The analysis of \cite{Riley:2021pdl} using the NICER and XMM data for PSR J0740+6620 gives $12.39^{+1.30}_{-0.98}$ km.    These analyses lead to an updated radius of 1.4$M_\odot$ neutron stars,\footnote{Here $R_{M/M_\odot}$ denotes the radius of a star of mass $M$.} $R_{1.4} = 12.45 \pm 0.65 $ km \citep{Miller:2021qha} and $12.33^{+0.76}_{-0.81}$ km \citep{Riley:2021pdl} to $12.18^{+0.56}_{-0.79}$ km \citep{Raaijmakers:2021uju}.

 Our aim in this paper is to delineate what the NICER measurements reveal about the microscopic properties of dense matter at baryon densities beyond nuclear saturation density $n_0$, and to translate this understanding into a physically motivated new equation of state of neutron star matter, QHC21. A key observation is that the effective lack of change of the radius as the mass increases from 1.4 to 2.08 $M_\odot$ strongly points to a very rapid growth of pressure with density in the range $\sim$ 2-4 $n_0$, a result confirmed in the equations of state of \cite{Miller:2021qha} used to fit the data.   Such behavior is in contrast to the gradual decrease of radii of neutron stars of masses well below 1 $M_\odot$ as the mass increases, a consequence of the pressure in nuclear matter not strongly resisting the increasing gravity.   The  lack of significant variation of the radii from $1.4M_\odot$ to $\sim 2.1M_\odot$ rules out substantial softening of matter (as would result from a strong first order phase transition) between 2-3$n_0$ and 4-5$n_0$ \citep{Drischler:2020fvz};  the radical shrinkage in the radius such softening would induce is incompatible with the new NICER data.

     A second key observation is that 
     the radius of a 1.4$M_\odot$ neutron star inferred from NICER measurements,
      for which the central baryon density is  $\sim$ 2-3$n_0$, indicates that matter in the nuclear regime has greater pressure than predicted earlier by microscopic many-body variational calculations of nuclear matter, as in the Akmal-Pandharipande-Ravenhall (APR)  and Togashi\footnote{The Togashi equation of state is based on precision two-body nuclear forces and empirical three-body nuclear forces, computed using variational techniques for nuclear many-body problems, as in APR the equation of state;  the same nuclear forces are used consistently in the nuclear liquid and crust equations of state. } 
equations of state \citep{Akmal:1998cf,Togashi:2017mjp}.   
The radius based on the NICER analyses is $\sim$ 0.7-0.8 km  
larger than predicted using earlier equations of state based in the nuclear regime on variational calculations -- for example, in QHC19  \citep{Baym:2019iky},  $R_{1.4} \simeq 11.6 $ km.  Since the radius of a $1.4M_\odot$ neutron star  is essentially determined by the equation of state up to densities $\sim 2 n_0$,
a larger radius requires greater stiffness here.
  Similarly, $2.1M_\odot$ neutron stars were predicted in QHC19 to have radii $\simeq$ 11.3-11.5 km.   Indeed, the NICER data provide the first observational test of equation of states for neutron star matter obtained microscopically  \citep{Akmal:1998cf,Togashi:2017mjp,Lonardoni:2019ypg,Drischler:2020fvz}.  
 
   The rapid pressure rise in the range $\sim$ 2-4 $n_0$ indicates the importance of quark degrees of freedom in this regime. Such a rapid rise does not naturally occur in descriptions of dense matter treating nucleons as elementary particles; rather, in models with realistic two- and three-nuclear forces \citep{Akmal:1998cf,Togashi:2017mjp}, the sound velocity gently rises without stop, eventually exceeding the causal limit at  densities $\gtrsim 5 n_0$. 
   The gentle growth in sound velocity can be also seen in relativistic mean field models of hadrons (e.g., \cite{Steiner:2012rk,DD2}).  
    
  Because nucleons are composed of quarks, increasing density causes the quark states at low momenta to become filled and triggers a rapid rise in pressure.  This process can be understood, initially, in terms of nucleons being pushed into the relativistic regime \citep{McLerran:2018hbz}, or more generally in terms of higher momentum quark states forming a Fermi sea, as \cite{Kojo:2021ugu} and \cite{Kojo:2021hqh} describe.  Such duality between relativistic nucleons and quark descriptions is a useful first picture of the domain intermediate between pure nucleonic and quark matter \citep{Baym:2019iky}.

     The NICER analysis of~\cite{Miller:2021qha} assumes the Togashi equation of state \citep{Togashi:2017mjp} for the crust up to baryon densities $n \lesssim 0.5n_0$, and parameterized equations of state -- a polytrope model, a model with a parameterized speed of sound, and a Gaussian model -- beyond that density.
On the other hand, the analyses of \cite{Riley:2021pdl,Raaijmakers:2021uju} use the equation of state from chiral effective field theory ($\chi$EFT, see below)  
 and polytrope or constant speed of sound models.\footnote{Recent analyses similar to NICER 
but with nuclear constraints from $\chi$EFT \citep{Pang:2021jta} yield somewhat lower estimates for the radii $R_{1.4 }=11.94^{+0.76}_{-0.87} $ km and $R_{2.08 }=11.96^{+0.80}_{-0.75} $ km than the NICER analyses.   
Previously the same group \citep{Dietrich:2020efo} gave $R_{1.4 }=11.75^{+0.86}_{-0.81} $ km without the data of PSR J0740+6620; the new NICER data leads to an increase in the estimate by $\simeq 0.2$ km, illustrating how data for $2M_\odot$ neutron stars constrains
the properties of neutron stars of lower mass and the equation of state in the core.}

        Here we construct a new equation of state of neutron star matter, QHC21, to incorporate the messages of NICER into a microscopic, rather than phenomenological, approach which links the neutron star equations of state to the underlying physics in QCD.      
        Microscopic descriptions, once established, enable one to explore correlations among various microscopic quantities and matter composition, and reveal important observable consequences in, e.g., neutron star cooling, and neutrino emission from supernovae and neutron star mergers.
                
        We use the  chiral effective field theory ($\chi$EFT) nuclear equation of state \citep{Lonardoni:2019ypg,Drischler:2020fvz,Drischler:2021kxf}  at densities below $\sim 1.5 n_0$.   This  approach, based essentially on an expansion in the effective momentum transfers, can provide  control of the uncertainty.   As discussed by \cite{Drischler:2020fvz} (see their Fig.~1), the uncertainty in the pressure within $\chi$EFT is small at $n_0$, 
 and grows to $\sim$  25\% at $\sim 1.4 n_0$;  the uncertainties at 2$n_0$ are too large to continue to use this approach at such density.

   We focus on nuclear matter calculations based on microscopic forces.    While phenomenological 
equations of state can also reproduce nuclear matter properties at densities $\sim n_0$ (e.g., \cite{Serot:1984ey}), 
the microscopic approach has the advantages that calculations can be improved both by systematically including 
higher order corrections and by refining the low energy constants in the chiral Lagrangian, and critically that they 
enable one to differentiate uncertainties in calculations of microscopic processes from uncertainties in the underlying 
physics, crucial in determining the importance of new beyond-nuclear physics towards high density.   On the other 
hand, mean-field Lagrangian approaches are flexible in adding new degrees of freedom and in fitting new data, but 
they have difficulties in finding good rationales to truncate large powers of field variables \citep{weinberg1995quantum}, 
which can significantly impact predicted properties of matter.

\begin{figure}[bt]
\vspace{-0.2cm}
\begin{center}	
	\includegraphics[width=8.8cm]{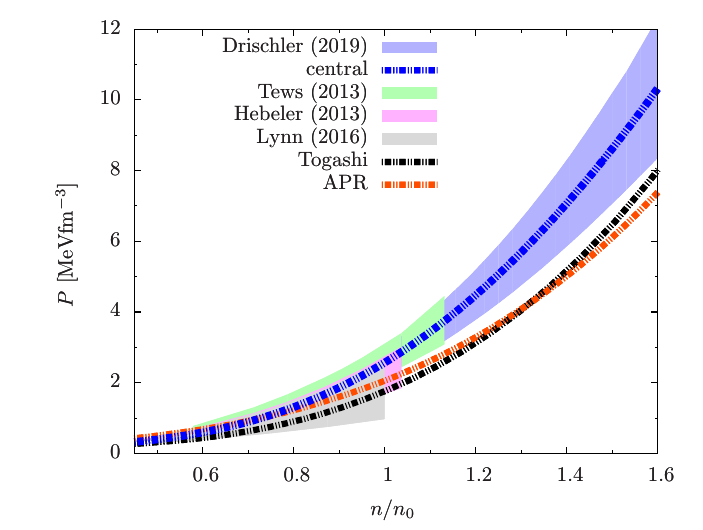}
\caption{ \footnotesize{ 
Nuclear equations of state for neutron star matter in beta equilibrium based on $\chi$EFT (collected in \cite{Raaijmakers:2021uju}) and variational methods based on 
phenomenological nuclear forces \citep{Akmal:1998cf,Togashi:2017mjp}.  The
``central" $\chi$EFT curve corresponds to the average of the data in \cite{Drischler:2020fvz}.   
		} }
\end{center}		
		\vspace{+0.3cm}
\label{fig:ChEFT_vs_Togashi}		
\end{figure}

     Figure~\ref{fig:ChEFT_vs_Togashi} shows the  range of the pressure in nuclear equations of state vs.~baryon density in $\chi$EFT approaches (labeled Lynn, Hebeler, Tews and Drischler)  and variational approaches (labeled APR and Togashi). The variational approaches are significantly softer, as consequence we expect of approximations made in constructing the variational solutions.   To allow for stiffer matter, we focus on the nuclear equations of state constructed by  \cite{Drischler:2020fvz} based on beyond-leading-order (N$^3$LO) $\chi$EFT.   Specifically we take the central value of the N$^3$LO  $\chi$EFT pressure  at given density.  This pressure, larger than in the Togashi equation of state,  by $\sim$ 30-40\% around $n_0$, leads to larger radii.   
 We use the  $\chi$EFT nuclear equation of state at densities below $\sim 1.5 n_0$.  We also indicate results using the Togashi equation of state,  thus covering the uncertainty range of microscopic nuclear calculations at lower densities.

    As discussed above, the rapid rise in pressure indicated by the NICER data well motivates the consideration of 
     the onset of higher momentum degrees of freedom around 2-3 $n_0$, and a breakdown in an approach based on interactions of nucleonic degrees of freedom.  Certainly at higher densities the degrees of freedom become those of strongly interacting quarks.  To facilitate the transition to quark degrees of freedom, we allow a transition region between nucleonic and quark degrees of freedom within the Quark-Hadron Crossover (QHC) framework \citep{Baym:2017whm,Baym:2019iky}.   
     We explore a greater range of equations of state above nuclear matter density that are more readily accessible by non-nucleonic descriptions, 
     accommodating an earlier breakdown of the nuclear description, and permitting possibly stiffer interpolated equations of state,
by using explicit nuclear degrees of freedom only up to $n \sim 1.5n_0$.        
    
    We use quark matter equations of state, calculated as calculated within the Nambu-Jona-Lasinio (NJL) model, now starting at a lower density, $n\gtrsim 3.5 n_0$, than $ \sim 5n_0$ where baryons  of vacuum radius 0.8 fm begin to overlap.\footnote{QHC19 previously used the Togashi nuclear equation of state at $n \le 2n_0$, and quark matter equations of state calculated 
within the NJL model for $n\gtrsim 5n_0$.}   
This modification allows for the possibility that quark exchanges in baryon interactions drive earlier overlap of baryons and hence earlier onset of quark matter.
In the transition region between  nuclear and quark degrees of freedom we carry out a smooth and highly constrained interpolation.\footnote{Although theoretical calculations in the crossover region are difficult, possibly involving baryon-like and quark-like degrees of freedom at the same time \citep{McLerran:2018hbz,Fukushima:2020cmk,Kojo:2021ugu}, the  
physics in this regime is highly constrained by having to match the nuclear and quark matter equations of state at the boundaries of the crossover region.  The constraints that the sound velocity not exceed the speed of light, and that the matter in the interpolation region be thermodynamically stable further constrain possible interpolations.  The interpolating curves reveal the general trends \citep{Masuda:2012kf,Masuda:2012ed,Masuda:2015kha,Kojo:2014rca,Annala:2019puf}.}   
 At high densities, the NJL model provides  the best available equation of state taking into account the restoration of chiral symmetry as well as color superconductivity in strongly interacting quark matter.    As we see in detail below, the new equation of state well accommodates the NICER data, and indicates the extent to which the central density of PSR J0740+6620 is high enough to contain quark matter.

\section{Physics of QHC21}
\label{sec:QHC}

  We detail here the construction of QHC21.   We specify the thermodynamics in terms of the pressure $P$ as a function of baryon chemical potential $\mu_{_B}$.   Nuclear equations of state are used up to a density $n^L$, and quark equations of state above a density $n^U$ ($> n^L$).  In the crossover between these regimes, we smoothly interpolate the pressure, the baryon density $n=\partial P/\partial \mu_{_B}$, and the susceptibility $\chi_{_B} = \partial^2 P/\partial \mu_{_B}^2$ at the matching boundaries.  
Thermodynamic stability imposes the requirement that the interpolated pressure leads to non-negative $\chi_{_B}$, 
and causality that the adiabatic speed of sound $c_s = \sqrt{ \partial P/\partial \varepsilon \,}$ does not exceed the speed of light, $c$.     
Here $\varepsilon = -P + \mu_{_B} n$ is the energy density.

 In QHC21 we consider primarily the 
$\chi$EFT equation of state in the nucleonic regime, but also note results using the Togashi equation of state there (labeled QHC21T).
To describe quark matter we employ the Nambu--Jona-Lasinio (NJL) model.  
 Although the NJL model does not include confining effects it is a good framework in which to capture dynamical effects, e.g., chiral symmetry breaking, concisely \citep{Hatsuda:1994pi}, and in addition 
 it reproduces quantities that are not very sensitive to confinement 
such as effective quark masses and low energy constants in the chiral effective Lagrangian for pions.   QHC thus incorporates general features of quark matter by exploring a wide range of the NJL model parameters consistent with neutron star observables.
The coupling constant $G$ and the ultraviolet cutoff $\Lambda$ of the NJL model, specified in \cite{Baym:2017whm}, are the relevant parameters describing chiral symmetry breaking. In addition, we include the short range density-density repulsion in QCD \citep{Kunihiro:1991qu,Song:2019qoh}, of strength measured by a coupling constant $g_{_{\rm V}}$, and an attraction between quarks, proportional to a coupling $H$, which governs the color superconductivity pairing gap.\footnote{For simplicity we assume that the effective coupling constants $g_{_{\rm V}}$ and $H$ are independent of density.  However, one expects from QCD that $g_{_{\rm V}}$ falls with increasing density \citep{Song:2019qoh}; see Appendix \ref{conformal_limit}.   The quark pairing interaction, $H$, should behave similarly.}

 The quark equations of state, parametrized by given $(g_{_{\rm V}}, H)$, and  the nuclear equation of state
  are first interpolated  in a given range $(n^L, n^U)$ by using fifth order interpolating polynomials in $\mu_{_B}$, for which the coefficients are uniquely fixed by six boundary conditions.
       We then determine the ranges of $(g_{_{\rm V}}, H)$ that are consistent with stability, causality, and the existence of the known high mass neutron stars.   This procedure predicts $M$ vs.~$R$, thermodynamic quantities such as baryon density as a function of $\mu_{_B}$, as well as more microscopic quantities such as effective quark masses and diquark paring gaps.

\begin{table}[bt]
\caption{
\vspace{2mm}
The interpolation ranges $(n^L, n^U)$ in units of $n_0$, and the interaction parameter sets $(g_{_{\rm V}}, H)$ for QHC21 A$_\chi$-D$_\chi$, and QHC21T A$_T$-D$_T$.   Tabulated for each parameter set are the maximum mass, $M_{\rm max}$, in units of $M_\odot$, radii $R$, in km, and central baryon densities, $n^c$, in units of $n_0$, of neutron stars of masses 1.4, 2.08 $M_\odot$, and $M_{\rm max}$.
}

\begin{center}
{\renewcommand\arraystretch{1.4}
\vspace{5mm}
\begin{tabular}{|c|cccc|cccc|}
\hline
& \multicolumn{4}{|c|}{ {\bf QHC21 } } &  \multicolumn{4}{c|}{ {\bf QHC21T} } \\
& A$_{\chi}$ & B$_{\chi}$ & C$_{\chi}$ & D$_{\chi}$ & A$_T$ & B$_T$ & C$_T$ & D$_T$  \\
\hline \hline
$(n^L, n^U) $ & \multicolumn{4}{|c|}{ (1.5, 3.5) $n_0$ } &  \multicolumn{4}{c|}{(1.5, 3.5) $n_0$ }  \\ 
\hline 
$g_{_{\rm V}}/G$ & 1.0 & 1.1 & 1.2 & 1.3 &  0.9 & 1.0 & 1.1 & 1.2 \\
$H/G$ & 1.50 & 1.52 & 1.54 & 1.56 &  1.50 & 1.52 & 1.55 & 1.57 \\
$M_{\rm max}$ & 2.19 & 2.25 & 2.31 & 2.37 &  2.13 & 2.20 & 2.26 & 2.32 \\
\hline
$R_{1.4} $  & 12.4 & 12.4 & 12.4 & 12.3 &11.8&11.9&11.7& 11.8 \\
$R_{2.08} $ & 12.0 & 12.2 & 12.4 & 12.5 &  11.5 & 11.8 & 11.9 & 12.0 \\
$R_{\rm M_{max} }$ & 11.7 & 11.5 & 11.4 & 11.2 & 10.9 & 11.1 & 11.1 & 11.3 \\
\hline
$n^c|_{1.4 }$ & 2.6 & 2.5 & 2.5 & 2.5 &  2.8 & 2.7 & 2.8 & 2.8 \\
$n^{c}|_{2.08 } $ & 4.3 & 4.0 & 3.5 & 3.3 &  4.9 & 4.2 & 4.0 & 3.6 \\
$n^c |_{M_{\rm max} } $ & 6.0 & 5.8 & 5.6 & 5.4 &  6.4 & 6.1 & 6.0 & 5.8 \\
\hline
\end{tabular}
}
\end{center}
\label{default}
\end{table}%

\

\section{ Neutron Star Structure }
\label{sec:NSS}

\begin{figure}[tb]
\vspace{-0.2cm}
\begin{center}	
\vspace{-0.2cm}
	\includegraphics[width=8.8cm]{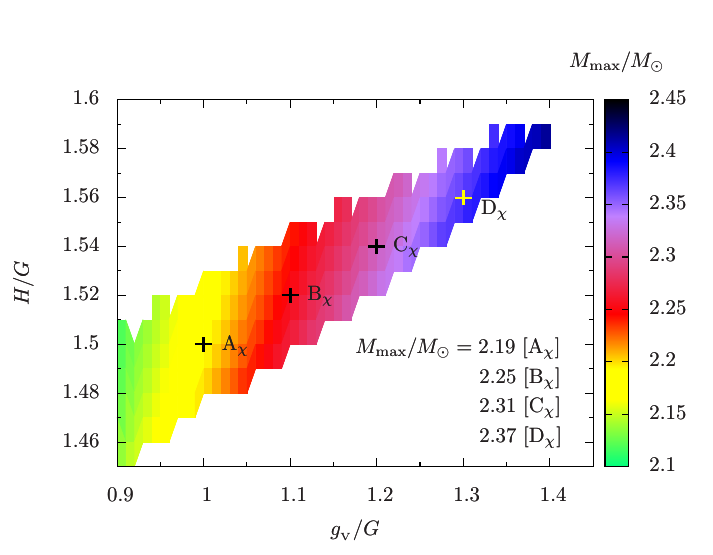}
\end{center}
\vspace{-1.7cm}
\begin{center}
	\includegraphics[width=8.8cm]{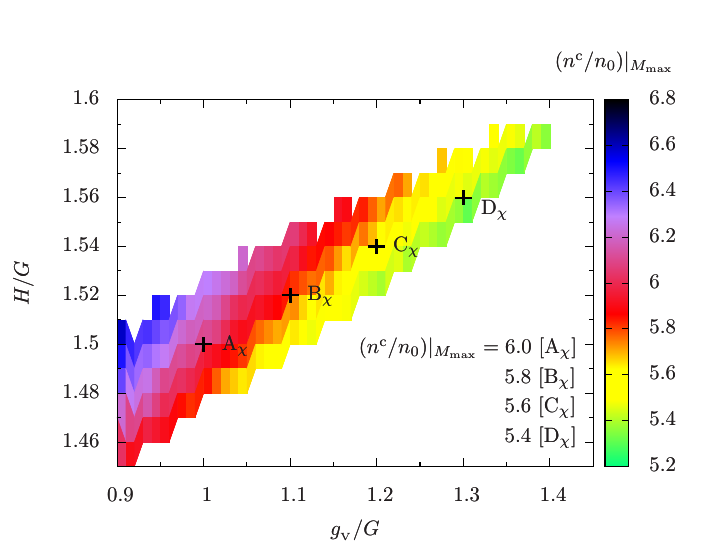}
\end{center}	
\vspace{-0.2cm}
\caption{ \footnotesize{ 
The maximum mass  $M_{\rm max}$  (upper panel) and the central density $n^c$ (lower panel) with the QHC21 equation of state.  The matching densities are $(n^L, n^U)/n_0 = (1.5, 3.5)$.
		} }
		\vspace{+0.3cm}
\label{fig:M-R_Mmax}		
\end{figure}

\begin{figure}[tb]
\vspace{-0.2cm}
\begin{center}	
\vspace{-0.2cm}
	\includegraphics[width=8.8cm]{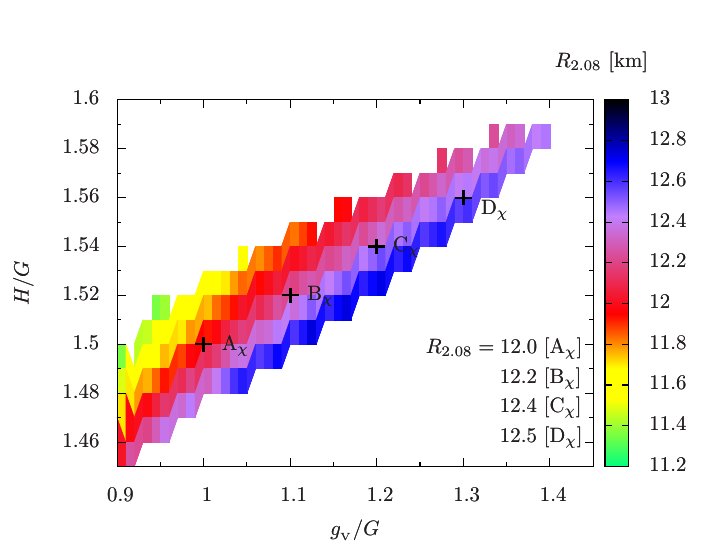}
\end{center}
\vspace{-1.7cm}
\begin{center}
	\includegraphics[width=8.8cm]{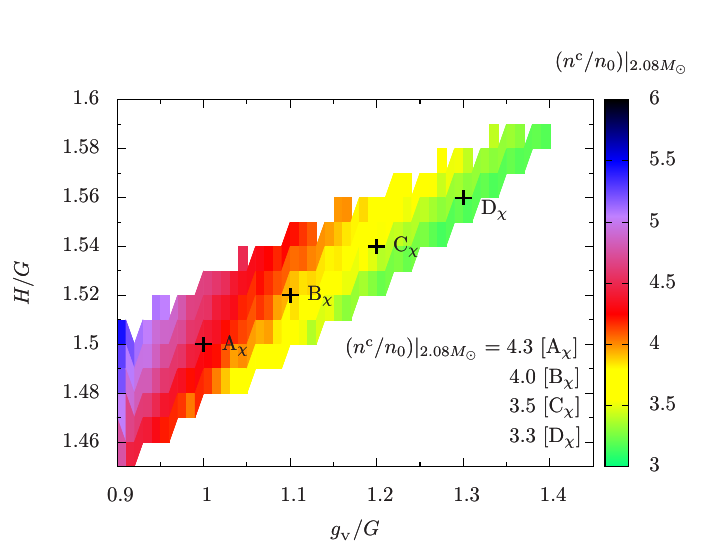}
\end{center}	
\vspace{-0.2cm}
\caption{ \footnotesize{
The radius (upper panel) and the central density  (lower panel) at $2.08M_\odot$,  in the ($g_{_{\rm V}},H$) plane with the QHC21 equation of state.
		} }
		\vspace{0.3cm}
\label{fig:M-R_M208}		
\end{figure}

\begin{figure}[tb]
\vspace{-0.2cm}
\begin{center}	
\vspace{-0.2cm}
	\includegraphics[width=8.8cm]{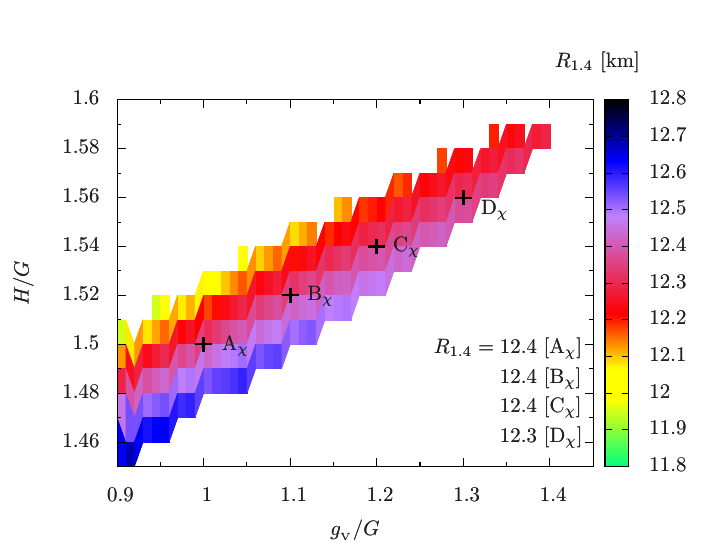}
\end{center}
\vspace{-1.7cm}
\begin{center}	
	\includegraphics[width=8.8cm]{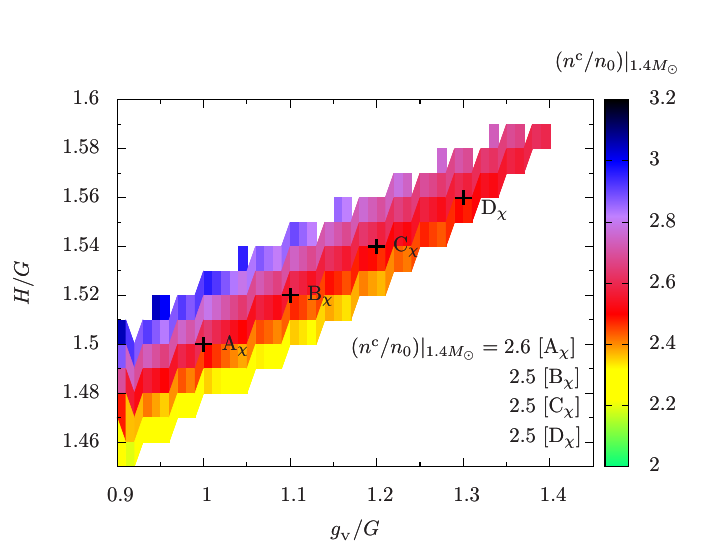}
\end{center}	
\vspace{-0.2cm}
\caption{ \footnotesize{ 
The radius (upper panel) and the central density (lower panel) at $1.4M_\odot$ for the same equations of state as in Figs.~\ref{fig:M-R_Mmax} and \ref{fig:M-R_M208}. 
		} }
		\vspace{0.3cm}
\label{fig:M-R_M140}		
\end{figure}

      Figures~\ref{fig:M-R_Mmax}-\ref{fig:M-R_M140} display, for $(n^L, n^U)/n_0 = (1.5, 3.5)$, the results of QHC21 
      in the $(g_{_{\rm V} }, H)$ plane for the maximum mass and the corresponding
central density, as well as the radius and central density of 2.08 $M_\odot$ and 1.4 $M_\odot$ stars.     We first determine the domain of $(g_{_{\rm V}}, H)$ that leads to equations of state satisfying the causality condition and that are consistent with the lower bound for the maximum mass, $M_{\rm max}/M_\odot \ge 2.08 \pm 0.07$;  these allowed regions are shown as colored in Figs.~\ref{fig:M-R_Mmax}-\ref{fig:M-R_M140}.  As a guide we choose several characteristic values in the $(g_{_{\rm V}},H)$ plane, focussing on 
 $(g_{_{\rm V}}, H)/G = (1.0, 1.50)$ (set A$_\chi$), $(1.1, 1.52)$ (set B$_\chi$), $(1.2, 1.54)$ (set C$_\chi$), and $(1.3, 1.56)$ (set D$_\chi$).  
 We denote these parameter sets with a subscript $\chi$ to distinguish them from the sets A, B, C, and D in QHC19.
In addition Table~\ref{default} summarizes the parameters and results of both QHC21 and QHC21T (with parameter sets denoted by a subscript T).  
 In Appendix \ref{range_of_interpolation} we vary the range of interpolation, and find that the domains of $(g_{_{\rm V}}, H)$ can vary by 10-20\%, with no qualitative changes.

     The upper panel of Fig.~\ref{fig:M-R_Mmax} shows the maximum neutron star mass $M_{\rm max}$, in the color scale on the right, as a function of ($g_{_{\rm V}}, H$) measured in units of the NJL coupling $G$.   
 Smaller repulsion $g_{_{\rm V}}$ leads to smaller maximum masses;   
 the requirement $M_{\rm max} \ge 2.08 M_\odot$ is satisfied only for 
$g_{_{\rm V}}/G \gtrsim 0.84$, while $g_{_{\rm V}}/G \lesssim 0.74$ leads to $M_{\rm max}/M_\odot \lesssim 2.01$.
The largest allowed\footnote{
This mass is in practice a lower bound on the absolute maximum, since we use only a minimal interpolation between the nuclear and quark regimes.  
In addition, as seen in Appendix B, reducing $n^U$ with $n^L$ fixed can lead to larger maximum mass.
}
$M_{\rm max}/M_\odot$ is $\simeq 2.43$ with $(g_{_{\rm V}}, H) \simeq (1.42, 1.59)$.  

     The lower panel of Fig.~\ref{fig:M-R_Mmax} shows the central density, $n^c$, at $M_{\rm max}$.  A larger $M_{\rm max}$ allows a smaller $n^c$, consistent with the general rule that the stiffer the equation of state the larger the maximum mass and the smaller the central density at the maximum.  
The restriction $M_{\rm max}/M_\odot \ge 2.08$ imposes the condition $n^c/n_0 \lesssim 6.4$. At $M_{\rm max}/M_\odot \simeq 2.40$, $n^c/n_0 \simeq 5.3$.

    Figure~\ref{fig:M-R_M208} shows the dependence of the radius and the central density of $2.08M_\odot$ neutron stars on  $g_{_{\rm V}}$ and $H$.
Within the allowed band of $(g_{_{\rm V}}, H)$, the radii range from $\simeq 10.8$ km to $\simeq 12.9$ km. 
Radii $\lesssim 11$ km,  significantly smaller than $R_{1.4} \simeq 12$ km, appear for equations of state whose maximum mass barely reaches $2.08 M_\odot$, since  the radius in general shrinks radically near the maximum mass.    

   Apropos of whether the central density of PSR J0740+6620 with a mass $2.08 M_\odot$ is high enough to accommodate quark matter as characterized by overlaps of baryons,  we see in the lower panel of Fig.~\ref{fig:M-R_M208} that the central density is $\lesssim 5n_0$, slightly smaller than the density where baryons of radius $\sim 0.8$ fm begin to overlap,  
   and is $ \gtrsim 3n_0$, comparable to $n^U = 3.5n_0$ as adopted here. 
   What is certain is that the core density is considerably larger than the density where pure hadronic calculations are applicable.
  The equations of state at a density slightly below $n^U$ are strongly correlated with quark matter models at densities $ \simeq n^U$. 
 
\begin{figure}[tb]
\begin{center}	
\vspace{-0.2cm}
	\includegraphics[width=8.9cm]{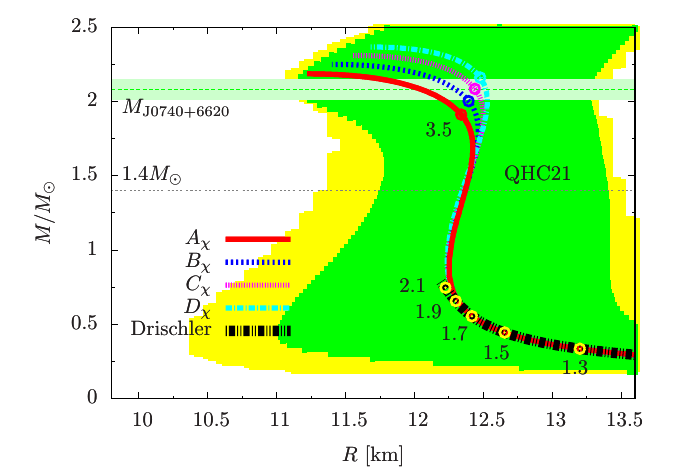}
	\includegraphics[width=8.9cm]{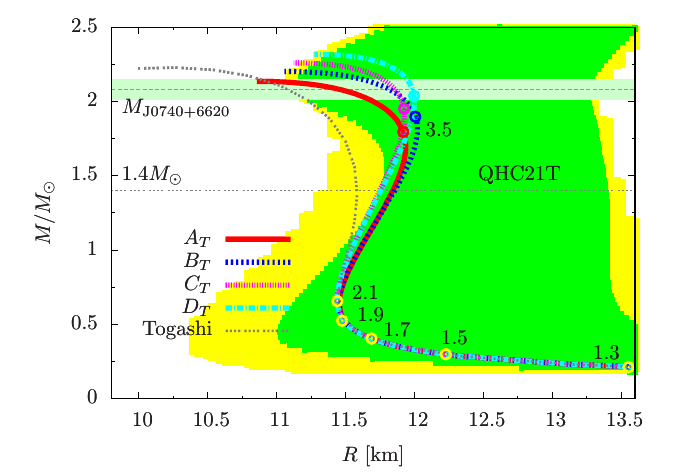}

\caption{ \footnotesize{
Upper panel: Mass vs.~radius for QHC21 (A$_\chi$-D$_\chi)$ and the ``central" $\chi$EFT equations of state \cite{Drischler:2020fvz}.
Lower panel: Mass vs.~radius for QHC21T (A$_T$-D$_T$) and pure Togashi equations of state. 
 In both figures,  the heavy dots indicate central densities $n^c = 1.3, 1.5, 1.7, 1.9, 2.1$ and 3.5 $n_0$. 		
The bands (95\% CL) correspond to ``PSR+GW+J0030+J0740'' in \cite{Legred:2021hdx} (yellow) and ``All Measurements (Gaussian Process)'' in \cite{Miller:2021qha} (green). }
}

			\vspace{0.2cm}
\label{fig:M-R_A-D}		
\end{center}
\end{figure}

 The central density, and thus the presence of quark matter, is sensitive to how
close the mass is to the maximum, $M_{\rm max}$, allowed by the equation 
of state.  This is because the radius shrinks rapidly by $\sim$ 0.5-1 km 
in a small mass interval $\sim 0.05 M_\odot$ below $M_{\rm max}$ as seen from  Fig. \ref{fig:M-R_A-D}.
The radius $R_{2.08}$ being comparable to $R_{1.4}$  indicates that
$2.08 M_\odot $ is still close to the lower edge of the above interval.
 Hence, we expect that $M_{\rm max}$ is larger than $2.08M_\odot$ by at least 0.05$M_\odot$.  
For $(g_{_{\rm V}}, H)$ satisfying this condition, $n^c|_{2.08} \lesssim 5n_0$.

     The range of radii of $1.4 M_\odot$ neutron stars, $\simeq 11.8$-$12.7$ km, is consistent with the NICER result $R_{1.4} = 12.4 \pm 0.6$ km.  The corresponding central density is 2.1-3.2$n_0$.   These radii are $\sim 0.3$-$1.2$ km larger than in QHC19 ($R_{1.4} \simeq 11.5$ km, with  $(n^L, n^U)/n_0 = (2.0, 5.0)$), a consequence primarily of using the stiffer $\chi$EFT nuclear equation of state.  The Togashi equation of state, even if one allows more freedom for the interpolated pressure in the density range 1.5-2.0 $n_0$, leads to radii $\simeq$ 11.6-12.2 km, on the smaller side of the NICER band.   
          Subsequent NICER determinations of the radii of neutron stars with intermediate masses would provide a further observational test of the inferred rapid pressure rise in the equation of state.
        The estimate of $R_{1.4}$ can be further improved by additional gravitational wave data from neutron star mergers.
          The tidal deformability extracted from the GW170817 event tends to favor a smaller $R_{1.4}$ than extracted from the NICER data alone, and this trend affects the NICER inference of $R_{1.4} = 12.4 \pm 0.6$ km used in the present paper.
          Provided the $\chi$EFT results stay largely unchanged in the future, 
          a larger $R_{1.4} \,(\gtrsim 12.4\,{\rm km})$ would require even more rapid stiffening than in QHC21, while a smaller one would shift the onset of rapid stiffening to higher densities.

\begin{figure}[tb]
\begin{center}	
\vspace{-0.2cm}
	\includegraphics[width=8.9cm]{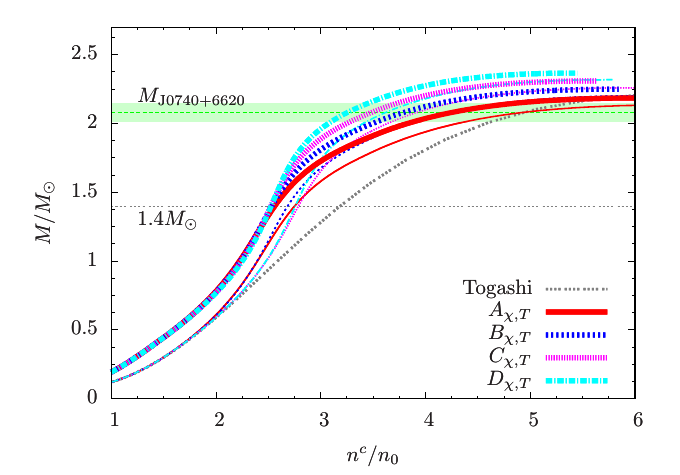}
\caption{ \footnotesize{
Mass vs. central density for QHC21 A$_\chi$-D$_\chi$ (bold lines),  QHC21T A$_T$-D$_T$ (thin lines), and the pure Togashi equations of state. 
} }
\label{fig:M-nb_A-D}		
\end{center}
\end{figure}

\begin{figure}[tb]
\begin{center}	
\vspace{-0.2cm}
	\includegraphics[width=8.9cm]{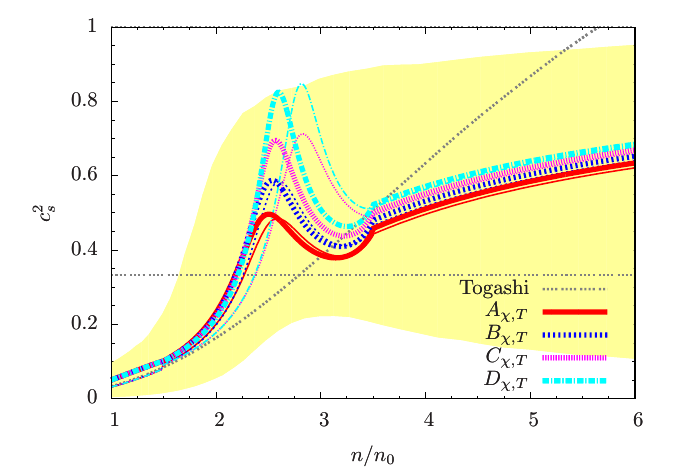}
\caption{ \footnotesize{
The square of the sound speed, $c_s^2$, in units of $c^2$, as function of baryon density $n$ for the QHC21 A$_\chi$-D$_\chi$ (bold lines), QHC21T A$_T$-D$_T$ (thin lines), and pure Togashi equations of state.   The conformal limit $c_s^2=c^2/3$ is shown as a guide. 
The band (95\%CL) shows the constraint from ``PSR+GW+J0030+J0740'' deduced by \cite{Legred:2021hdx}).
} }
			\vspace{-0.2cm}
\label{fig:cs2-nb_A-D}		
\end{center}
\end{figure}

    Having described the overall trend of the masses, radii, and central densities in the parameter space, we look more closely at results for the parameter sets A$_\chi$-D$_\chi$ for QHC21.   The $M$-$R$ relation for these parameters is shown in the upper panel of Fig.~\ref{fig:M-R_A-D}; the lower panel shows the corresponding results for QHC21T for the parameter sets A$_T$-D$_T$.  
The numbers next to the heavy dots indicate the central density of the star, in units of $n_0$, for the given mass.    
As we have matched interpolating functions with nuclear equations of state up to second derivatives at $n=1.5n_0$, 
the deviation of the $M$-$R$ curves from what the pure, i.e., without quark matter, $\chi$EFT or Togashi equations of state would give appears only after $n$ increases considerably beyond the matching point $n=1.5n_0$. 
(This trend can also be seen in the speed of sound in Fig.~\ref{fig:cs2-nb_A-D}.) 
The central $\chi$EFT equation of state in Fig.~\ref{fig:ChEFT_vs_Togashi}, which is available only for $n \lesssim 2.2n_0$, predicts overall larger radii for low core densities, leading to better agreement of QHC21 with the NICER data.
As we see in  Fig.~\ref{fig:M-nb_A-D}, in QHC21 the equations of state become stiffer than QHC21T and pure Togashi at $\sim 2n_0$, producing $1.4M_\odot$ neutron stars at a lower central density than with Togashi alone.

    An important feature of the QHC21 equation of state is the behavior of the sound velocity.\footnote{We note that the density dependence of the NJL couplings can change $c_s^2$ in the quark matter region. The couplings should decrease at very large density so that the equations of state are dominated by the quark kinetic energy and $c_s^2$ approaches $1/3$, as it should in the high density limit.}  
As seen in Fig.~\ref{fig:cs2-nb_A-D}, $c_s^2$ exceeds the conformal limit $c_s^2=c^2/3$ at $n \simeq$ 2.1-2.3$n_0$,  
a considerably lower density than $\simeq 2.8n_0$ with Togashi alone, and also less than the $\sim 2.6n_0$ in QHC19. 
The peaks in $c_s^2$ for the QHC21 set A$_\chi$-D$_\chi$ or A$_T$-D$_T$ are at $\simeq$ 2.1- 3.0 $n_0$, a considerably lower density than  in QHC19 they are in the range $\simeq 3.0$-$4.0n_0$.  

   Purely nucleonic models achieve stiff equations of state by having repulsive three- or more-body forces at short distance.  Including only few-body forces, such models constrained by low density experiments lead to slow growth in $c_s^2$, as seen in the Togashi curve in Fig.~\ref{fig:cs2-nb_A-D}.  Quicker growth requires more-body forces whose dominance would invalidate using nucleons as effective degrees of freedom.  The rapid growth of $c_s^2$ found in QHC21 is unlikely to be achieved by nucleonic models, differentiating QHC from nuclear models.

   The peak in the sound velocity is a novel feature of the crossover at finite baryon density, 
 as originally pointed out 
phenomenologically by \cite{Masuda:2012ed} and recently analyzed theoretically by \cite{McLerran:2018hbz,Kojo:2021ugu,Kojo:2021hqh} 
in terms of the quark substructure of baryons and quark Pauli blocking.  In the crossover domain, quark Pauli blocking puts kinematic constraints on baryons pushing them to be relativistic.  In quark descriptions, relativistic quarks start to contribute  directly to the thermodynamic pressure even before the quark Fermi sea is established.

 The theoretical estimates \cite{Kojo:2021ugu,Kojo:2021hqh} 
  suggest that quark Pauli blocking effects set in around $\sim 1$-$3n_0$, radically increasing the non-relativistic pressure $\sim n^{5/3}/m_{_N} $ (where $m_{_N}$  is the nucleon mass) to the relativistic behavior $\sim N_c n^{4/3} $, where $N_c(=3)$ counts the number of quarks in a baryon. 
The corresponding changes in the energy density are modest; $ \sim m_{_N} n$ in non-relativistic regime and $\sim N_c n^{4/3}$ in relativistic regime are comparable in the crossover domain.  Such a rapid growth in pressure with modest change in energy density leads to rapid enhancement of $c_s^2$.   As this transient regime comes to a close, both the pressure and energy density scale as in relativistic regime and hence $c_s^2$ eventually relaxes to $1/3$, leaving a sound velocity peak in the crossover domain.

  A sound velocity peak of relativistic origin does not exist in a hot QCD plasma \citep{HotQCD:2014kol},  where the speed of sound develops a dip instead of a peak in the crossover region.  The finite temperature transition from a hadron resonance gas to a quark gluon plasma is largely driven by non-relativistic resonances which are energetically disfavored but are important due to their large entropies.  Such entropic effects are absent in cold dense matter.

\begin{figure}[htb]
\begin{center}	
\vspace{-0.2cm}
	\includegraphics[width=8.8cm]{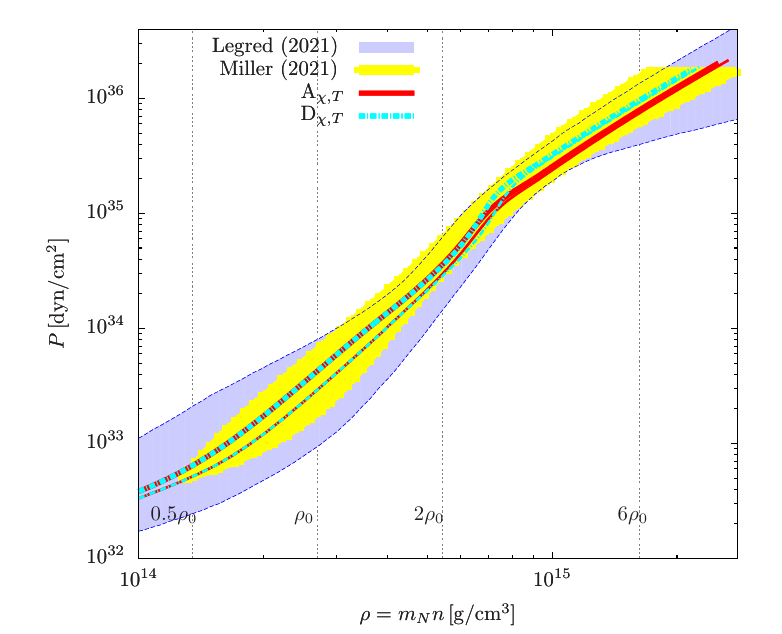}
\caption{ \footnotesize{
Pressure in QHC21(A$_{\chi,T}$, D$_{\chi,T}$) compared with the limiting constraints deduced by the NICER analyses, as a function of baryon rest mass density $\rho =m_N n$.  
The vertical lines,  $\rho$ = 0.5, 1.0, 2.0, and 6.0 $m_N n_0$, are shown as guides. 
The bands correspond to ``PSR+GW+J0030+J0740'' in \cite{Legred:2021hdx} (blue) and ``All Measurements (Gaussian Process)'' in \cite{Miller:2021qha} (yellow). 
Both bands are 95\% CL. 
The band appears narrow in the region $\rho \sim 10^{15}$ gm/cm$^3$ only as a consequence of plot being logarithmic.
}}			
\vspace{+0.2cm}
\label{fig:Prho}		
\end{center}
\end{figure}

\begin{figure}[b]
\begin{center}	
\vspace{-0.0cm}
	\includegraphics[width=8.9cm]{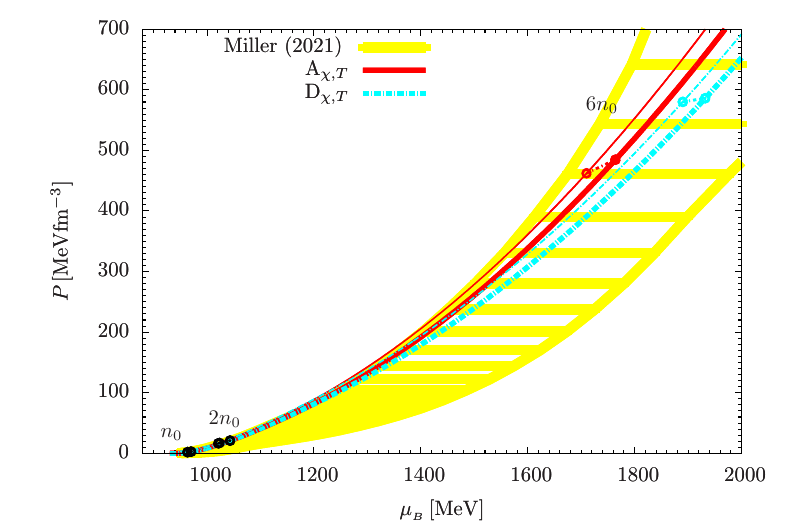}
\caption{ \footnotesize{
Pressure as a function of baryon chemical potential $\mu_{_B}$.  
The heavy dots on the QHC21 curves indicate densities $n = 1, 2$ and $6n_0$. 
The band (95\% CL) shows the equation of state bound ``All Measurements (Gaussian Process)'' in \cite{Miller:2021qha} . 
} }
\label{fig:e-P}		
\end{center}
\end{figure}

  Finally, we compare the pressure in QHC21 with constraints deduced by the NICER analyses.  Figure~\ref{fig:Prho} shows the pressure in QHC21 as a function of the baryon mass density, $\rho = m_{_N} n$, where $m_{_N} = 1.67\times 10^{-24}$ g is the nucleon mass.   Then Fig.~\ref{fig:e-P} compares the pressure as a function of baryon chemical potential $\mu_{_B}$, with the constraint ``All Measurements (Gaussian Process)'' in \cite{Miller:2021qha} 
  at a 95\%CL.\footnote{
  The error bands in Fig.~~\ref{fig:Prho} are primarily at larger $P$ at given $n$ than the QHC21 curves, while in Fig.~\ref{fig:e-P} they occur for smaller $P$ at given $\mu_{_B}$.  This reversal is a consequence of thermodynamics.  For example, if the energy density, $\varepsilon$, is scaled by a factor $1+\eta$ at fixed $n$, then $P$ and $\mu_{_B}$ are similarly scaled.  But at fixed $\mu_{_B}$, the scaling of $P$ has the opposite sign: $\delta P|_{\mu_{_B}} = - \eta \varepsilon $.}  The QHC21 equations of state are quite consistent with the NICER inferences.

\section{Discussion}
\label{sec:Discussion}

    Equations of state that exhibit a peak in the sound velocity occurring at the relevant intermediate densities robustly satisfy constraints from the NICER data.
      In the quark regime, matter can be stiffened by several physical effects (as discussed in greater detail in Appendix \ref{conformal_limit}).  First,  the vector repulsion between quarks, which allows quark matter to support neutron stars above $2M_\odot$, stiffens the matter in general, and produces sound velocities above the conformal limit, $c/\sqrt3$.  Furthermore, quark pairing correlations leading to color superconductivity \citep{Alford:2004pf}\footnote{Recent work suggests that an early onset of quark matter \citep{Annala:2019puf} or quark Pauli blocking effects in the transition regime  \citep{McLerran:2018hbz,Kojo:2021ugu} can stiffen matter at densities lower than predicted by typical hadronic models (see Fig.~2 in \cite{Kojo:2020krb} for possible scenarios), and that this stiffening at low density is consistent with observational constraints \citep{Tews:2018kmu}.  }
 or baryonic correlations in quarkyonic matter \citep{McLerran:2018hbz} can also stiffen the equation of state.\footnote{Underlying diquark pairing is the color magnetic interaction in QCD, which plays a crucial role as well in hadron spectroscopy \citep{DeRujula:1975qlm} and short distance baryon-baryon interactions \citep{Oka:1980ax,Park:2019bsz}.}


Within QHC21, the central density of PSR J0740+6620, $\lesssim 5n_0$, is well above densities where pure hadronic calculations are valid, and is entering the transition regime to strongly interacting quark matter. 
The cores of higher mass neutron stars, possibly to be discovered in the future, could reach the density region well beyond the peak in the sound velocity.  To describe such stars quantitatively, fully microscopic calculations of matter undergoing the transition from nucleonic to quark degrees of freedom will be needed.
   Nonetheless  the properties of quark matter strongly influence physics in the transition, possibly even slightly above saturation density.

Tables of the equations of state QHC21 for parameter sets A$_\chi$-D$_\chi$ and A$_T$-D$_T$ are available on the CompOSE archive at 
 \url{https://compose.obspm.fr/eos/236} through \url{https://compose.obspm.fr/eos/243}.
\section*{Acknowledgments}
G.B. would like to thank Cole Miller and Fred Lamb for informative discussions of the recent NICER results, including at the Aspen Center for Physics (supported by supported by National Science Foundation Grant No.~PHY-1607611).  
We thank Christian Drischler, Cole Miller, Geert Raajimakers, and Isaac Legred for kindly providing us with their data and instructions on using it.  
T.H. is supported by JSPS Grant No. JP18H05236 and JST CREST Grant No. JPMJCR19T2, and
T.K. by NSFC Grant No.~11875144 and by the Graduate Program on Physics for the Universe at Tohoku university.  

\widetext{}

\appendix

\

\section{A. The conformal limit for the sound velocity}\label{conformal_limit}

  In QCD at ultrahigh densities in the absence of significant length scales, the velocity approaches the conformal limit, $c_s =c/\sqrt 3$, from below; this limiting value is the conformal bound.    We focus in this Appendix on two questions:   The first is how the equation of state 
determines whether the sound velocity can exceed the conformal bound.\footnote{\cite{Tews:2018kmu}, also \cite{Greif:2018njt}, assuming the validity of nuclear calculations up to $n\sim 1.5n_0$ where $c_s^2$ is only $\sim$ 0.1 - 0.2,
point out that to have sufficient stiffening at higher density in order to reach neutron stars over two solar masses, it is necessary that $c_s^2$ exceed 1/3 there.   On the other hand, \cite{Annala:2019puf} construct stiff equations of state by allowing  $c_s^2$ to reach $\sim 1/3$ from below at $n = 1.1$-$1.5n_0$, a velocity well above the nuclear results; such early stiffening results in stiff equations of state that do not violate the conformal limit.
Such radical stiffening in the interval 1.1-1.5 $n_0$ could be tested in nuclear experiments.}
   In considering quark matter in neutron stars we are particularly interested in the domain 
where quark descriptions are natural but the matter is not dense enough for perturbative treatments (valid at $\gtrsim 40 n_0$).  While such a domain, above 5$n_0$, may not show up in the cores of neutron stars,  understanding $c_s$ there should give important constraints on the behavior of equations of state in the range of baryon density  $\sim 2$-$5 n_0$ which is only weakly constrained by perturbative results  alone.  
The second question  we discuss is the behavior of the sound velocity in QED, a theory that is not asymptotic free.
As we show, in the perturbative treatment of the massless electron gas, the sound velocity exceeds $c/\sqrt{3}$.
   We set $c=1$ here for simplicity. 

  The sound velocity involves a microscopic interplay between the kinetic energy density, $\varepsilon_{\rm kin}$ and the interaction energy, 
 $\varepsilon_{\rm int}$.  We first show schematically how the conformal bound can be violated in relativistic matter.  The quark number density, $n_{_{\rm Q}}$, is given in terms of the quark Fermi momentum, $p_{_{\rm F}}$, for equal mass quarks, by $n_{_{\rm Q}} = N_f  p_{_{\rm F}}^3/\pi^2$, with $N_f$ the number of quark flavors present; the kinetic energy density for massless quarks, is thus $\sim p_{_{\rm F}}^4 \sim
n_{_{\rm Q}}^{4/3}$.
Calculationally it is simplest to write the total energy density  $\varepsilon (n_{_{\rm Q}})$ in terms of $p_{_{\rm F}}$ as
\beq
  \varepsilon = a p_{_{\rm F}}^4 \left( 1+h(p_{_{\rm F}}) \right) , 
  \label{EA1}
\eeq
where $a= 3N_f/4 \pi^2$ and $h(p_{_{\rm F}})=\varepsilon_{\rm int}/\varepsilon_{\rm kin}$ is the interaction energy relative to the kinetic energy.  Then the pressure is 
\beq
  P = \frac{a}{3}p_{_{\rm F}}^4(1+h+ p_{_{\rm F}} h'),
  \label{PA1}
\eeq
where primes denote derivatives with respect to $p_{_{\rm F}}$.
Equations.~(\ref{EA1}) and (\ref{PA1}) imply
\beq
  \frac{\partial \varepsilon}{\partial p_{_{\rm F}}}  =   a p_{_{\rm F}}^3 \left( 4(1+h)+ p_{_{\rm F}} h' \right), \quad \ \ 
  \frac{\partial P}{\partial p_{_{\rm F}}}  = \frac{a}{3}p_{_{\rm F}}^3 \left( 4(1+h) + 6p_{_{\rm F}}h' + p_{_{\rm F}}^2 h'' \right) ,
\eeq
so that
\beq
    c_s^2  = \frac13 \left[1+\frac{5p_{_{\rm F}} h' + p_{_{\rm F}}^2 h''}{4(1+h)+ p_{_{\rm F}} h'}\right] . 
    \label{cs}
\eeq
In the absence of $h'$ and $h''$, the $h$ term does not affect the sound velocity. 

   Let us first consider $h$ having a power-law (PL) behavior, $h =b p_{_{\rm F}}^\zeta$, where $b$ is a constant.
 Then,  
\beq
    c_s^2({\rm PL})  =  \frac13 \left[1+\frac{(4+\zeta)\zeta h}{4+(4+\zeta) h }\right].
    \label{cs2}
\eeq
 We see that a repulsive interaction, $h >0$, with $\zeta > 0$ increases the sound velocity above the conformal bound.    An attractive interaction, $h < 0$,  can also stiffen the equation of state if $ \zeta<0$. Such density dependence is typical in correlation effects near the Fermi surface.   
 
 
 Pairing correlations in color superconductivity \citep{Alford:2004pf}
are in a regime with $\zeta \sim 2$ 
     to within the density dependence of the gap (which is not known well in the non-perturbative regime), and thus tend to increase the sound velocity.     
Baryonic correlations in quarkyonic matter \citep{McLerran:2018hbz} produce similar effects.
For example, the repulsive vector interaction between quarks in  the NJL model, $\sim g_{_V}n_Q^2$, has $\zeta = 2$, 
and
 in the three-flavor NJL model employed in the text, $h_{_{\rm NJL}} = (4/\pi^2) g_{_{\rm V}} p_{_{\rm F}}^2$.  Then   
   \beq
  c_s^2({\rm NJL})  =  \frac13 \left[1+\frac{6h_{_{\rm NJL}} }{2+3h_{_{\rm NJL}} }\right], 
    \label{cs-NJL}
\eeq
which always exceeds  1/3 and approaches to 1 in the high density limit.
   
On the other hand,  in the perturbative QCD (pQCD) regime at ultrahigh densities, beyond the range one actually encounters in neutron stars, $\zeta=0$, and an additional physical effect comes into play, the strong coupling constant $\alpha_s$ in dense matter runs with 
 $p_{_{\rm F}}$ (or equivalently the chemical potential) as  $ \sim 1/ \ln (p_{_{\rm F}}/\Lambda_{\rm QCD} )$ with $\Lambda_{\rm QCD}$ 
the QCD scale parameter ($\simeq$ 340 MeV for $N_f=3$).  To first order in $\alpha_s$, this dependence leads explicitly to 
\beq   
 \varepsilon_{_{\rm pQCD}} &=&\frac{3N_f}{4\pi^2}  p_{_{\rm F}}^4 \left(1+\frac{2\alpha_s}{3\pi}\right);
  \eeq
thus $h_{_{\rm pQCD}} = 2\alpha_s/3\pi$.  To lowest order 
\beq
 p_{_{\rm F}}   \frac{\partial \alpha_s}{\partial p_{_{\rm F}} } = -\frac{33-2N_f}{24\pi}\alpha_s^2,
  \label{alphaqcd}
\eeq
while $h''$ is of order $\alpha_s^3$.  From Eq.~(\ref{cs}) we then find for three flavors, 
\beq
    c_s^2 ({\rm pQCD})  \simeq  \frac13\left(1+\frac54 p_{_{\rm F}}   h' \right) =  \frac13 \,\left(1-\frac{5 (33-2N_f)}{144\pi^2}  \alpha_s^2 \right)
   \stackrel{N_f=3}{\longrightarrow}\frac13 \,\left(1-\frac{15}{16}  \left(\frac{\alpha_s}{\pi}\right)^2 \right),
   \label{cs-pQCD}
\eeq
approaching the asymptotic value 1/3 from below.  We see here that the asymptotic approach to the conformal limit is the result of a competition between the gluonic factor 33, which produces an approach from below, and the fermionic term $2N_f$, which produces an approach from above.   In QCD the gluonic term wins, and the approach is from below.

 Equation~(\ref{cs-NJL}), valid at intermediate densities relevant to neutron star interiors, and Eq.~(\ref{cs-pQCD}) for asymptotically high density,  are smoothly joined by  a density dependent vector interaction, $g_{_{\rm V}}^*$, as introduced in \cite{Song:2019qoh},
\beq
     g_{_{\rm V}}^*(p_{_{\rm F}})= \frac{4 \pi }{3} \frac{\alpha_s(p_{_{\rm F}})}{9 m_g^2 + 8 p_{_{\rm F}}^2}.
\eeq
Here  $m_g\simeq 400$ MeV is a non-perturbative scale proportional to $\Lambda_{\rm QCD}$, and 
the running coupling for $p_{_{\rm F}} < m_g$ is  frozen at  $\alpha_s(m_g)$ with $g_{_{\rm V}} \equiv {4\pi \alpha_s(m_g)}/27 m_g^2$.  
With this density dependent vector coupling,  $h =(4/\pi^2)g_{_{\rm V}}^* p_{_{\rm F}}^2$, which approaches $h_{_{\rm NJL}}$ at $p_{_{\rm F}} \ll m_g$ and 
 $h_{_{\rm pQCD}}$ at $p_{_{\rm F}} \gg m_g$.  The Fermi momentum at which ${c_s^*}^2 -1/3 $ changes sign from positive to negative is $\simeq 1.5 \Lambda_{\rm QCD}$, corresponding to a baryon density $n \sim 10 n_0$.

\begin{figure}[t]
  \begin{center}
   \includegraphics[width=8.0cm]{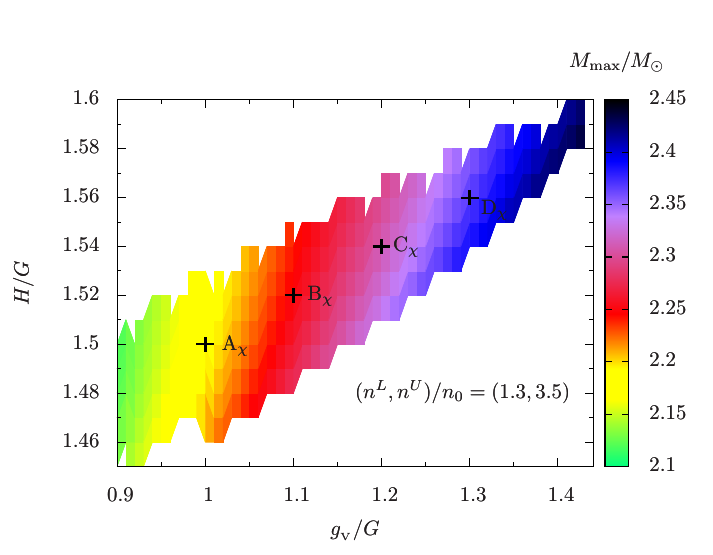}
   \includegraphics[width=8.0cm]{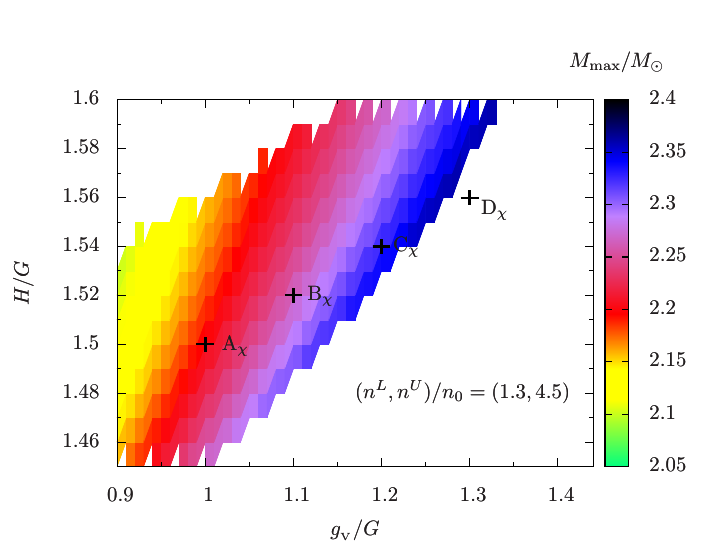} 
  \end{center}
  \vspace{-1.3cm}
 \begin{center}
   \includegraphics[width=8.0cm]{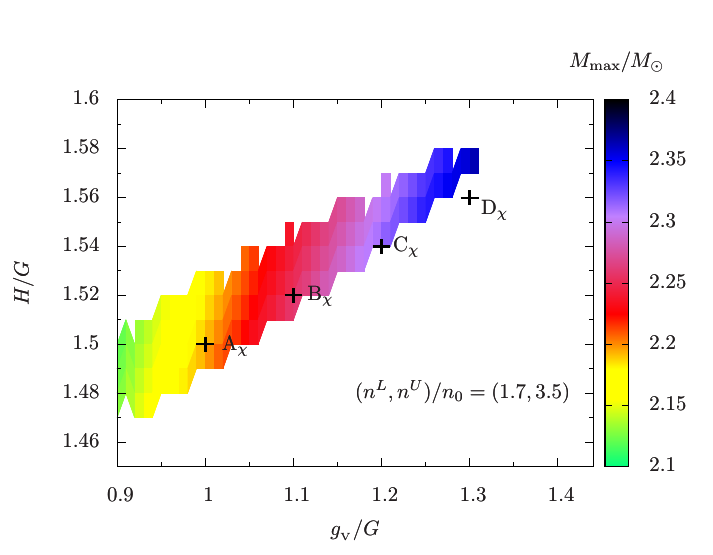}
   \includegraphics[width=8.0cm]{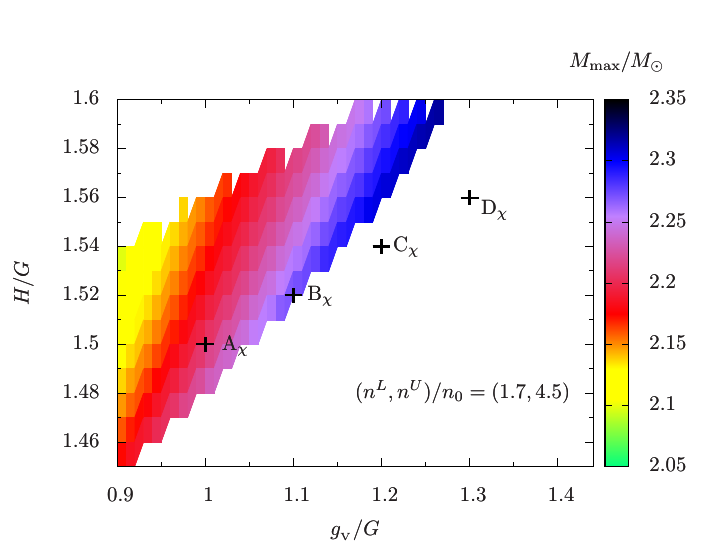} 
 \end{center}
  \caption{    \label{fig:gv-H-M-nL-nU} 
  \footnotesize{
The maximum mass  $M_{\rm max}$ as a function of $(g_{_{\rm V}}, H)$ for the four combinations of  matching densities $n^L$= (1.3, 1.7) $n_0 $ and $n^U$ = (3.5, 4.5) $n_0 $.   
The parameter sets $A_\chi$-$D_\chi$ in the main text are indicated only as guides.  
Outside the colored regions the system primarily violates thermodynamic stability.
}  }
\vspace{12pt}
\end{figure}

 Quantum electrodynamics, in contrast to QCD, has no ``charged" gluons,  
so that it is not asymptotically free and the sound velocity exceeds 1/3.
    To see this in detail, we note that the energy density in the relativistic limit to first order in $\alpha_e$ is
  \beq
    \varepsilon_{_{\rm QED}} &=& \frac{1}{4\pi^2}  p_{_{\rm F}}^4  \left(1+\frac{\alpha_e}{2\pi}\right),
   \eeq
so that $h= \alpha_e/2\pi$.  The evolution of $\alpha_e$ in QED is equivalent to dropping the 33, letting $N_f=1$ and 
 $\Lambda_{_{_{\rm QCD}}} \rightarrow m_e$ in $\alpha_s$, and multiplying by a factor 2 which arises from the algebraic difference of the interactions in QED and QCD. 
  Then, to lowest order,   we have 
$p_{_{\rm F}} \partial \alpha_e /\partial p_{_{\rm F}} = \alpha_e^2/6\pi$.  The resulting sound speed, to order $\alpha_e^2$, is given by
\beq
    c_s^2 ({\rm QED}) = \frac13 \,\left(1+\frac{5}{48} \left(\frac{\alpha_e}{\pi}\right)^2 \right),
\eeq
which exceeds 1/3 (and grows as the Fermi momentum increases).
In the low density non-relativistic limit, the sound speed of the electron gas is 
$c_s^2 ({\rm QED}) = (p_{_{\rm F}} /m_e)^2/3$, which is smaller than 1/3.

\section{B. Exploring other interpolation ranges}\label{range_of_interpolation}

   While we take the interpolation range $n^L=1.5 n_0$ and $n^U =3.5 n_0$ for our main results in the text, we examine in this Appendix the
dependence of the NJL parameters and the resulting mass-radius relations on the matching densities $n^L$ and $n^U$.  
We consider $n^L = (1.3, 1.7)  n_0$ and $n^U = (3.5, 4.5) n_0$ in the four possible combinations, continuing to use the
$\chi$EFT equation of state in the nuclear regime.  
Figure~\ref{fig:gv-H-M-nL-nU} shows the allowed ranges of $(g_{_{\rm V}}, H)$, 
with the parameter sets A$_\chi$-D$_\chi$ in the main text marked as guides.
As we can see, changing  the interpolation range modifies the domains of $(g_{_{\rm V}}, H)$ by only 10-20\%.

   Increasing the size of interpolating interval should in principle monotonically extend the window for $(g_{_{\rm V}},H)$, as well as allow a larger mass range.  
   For example, the results of $(n^L, n^U) =(1.3, 4.5) n_0$ should contain the results of $(1.7, 3.5) n_0$ as special cases.  
However in our modeling where we use simple polynomials which cannot cover all allowable curves, this is not the case.
We use simple structureless polynomials so as not to introduce exotic structures by hand, and yet the interpolation results in a highly nontrivial structure in the equation of state, e.g., the peak in sound velocity.   
Owing to our not exploring {\it all} possible interpolation curves, the maximum mass we find in a given interpolation window should be understood as a lower bound of  the maximum mass.

\begin{figure}[t!] 
 \begin{minipage}{0.33\hsize}
  \begin{center}
   \includegraphics[width=6.0cm]{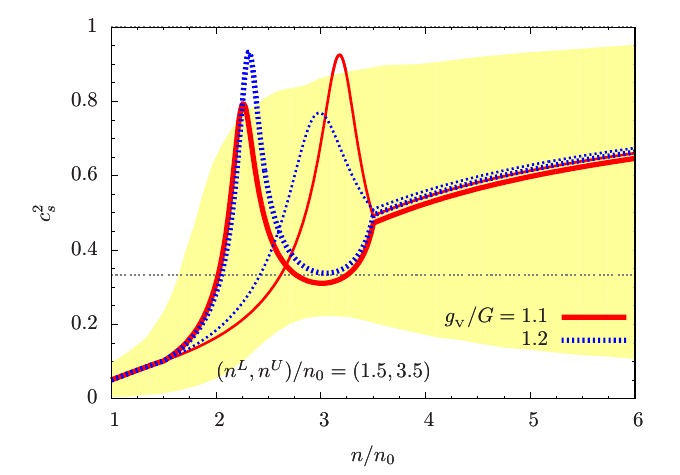}
   \includegraphics[width=6.0cm]{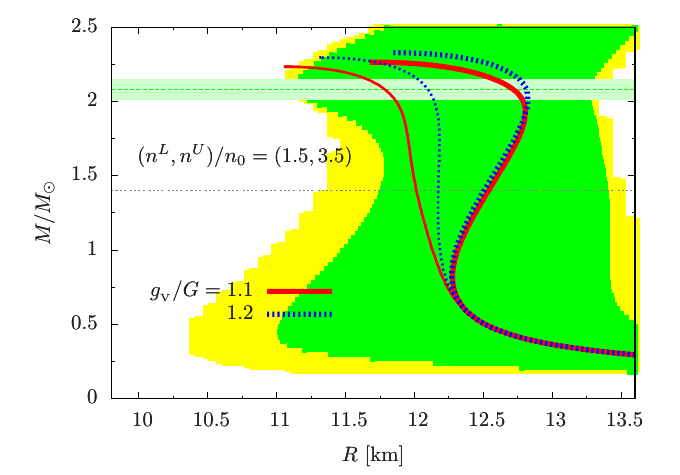}
   \includegraphics[width=6.0cm]{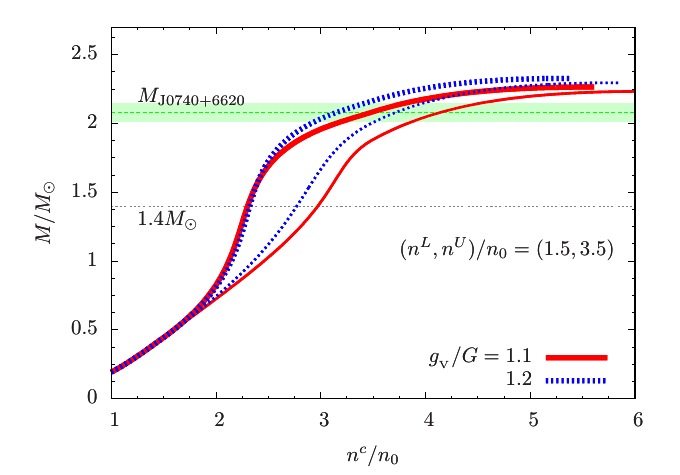} 
  \end{center}
 \end{minipage}
 \begin{minipage}{0.33\hsize}
 \begin{center}
   \includegraphics[width=6.0cm]{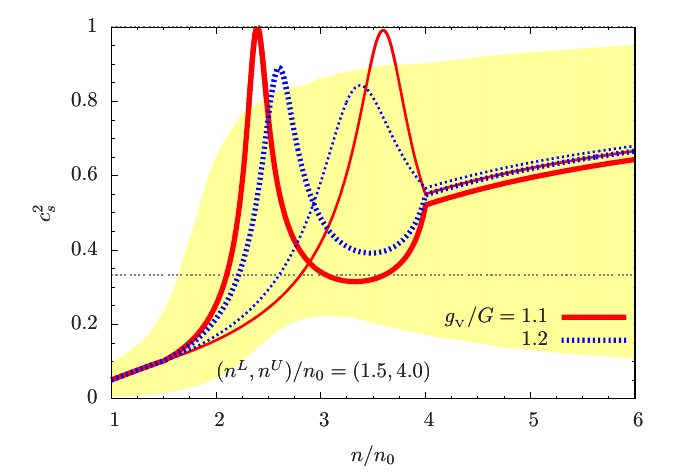}
   \includegraphics[width=6.0cm]{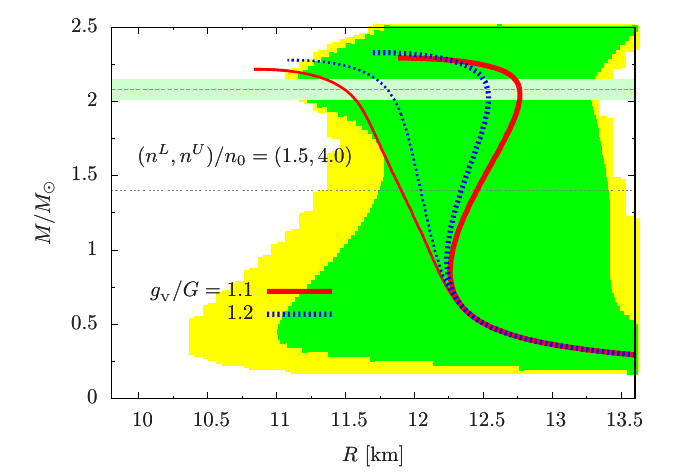}
   \includegraphics[width=6.0cm]{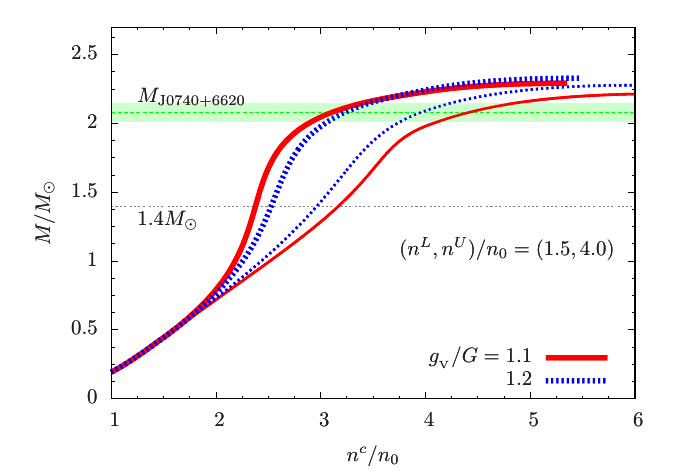} 
 \end{center}
 \end{minipage}
 \begin{minipage}{0.33\hsize}
 \begin{center}
   \includegraphics[width=6.0cm]{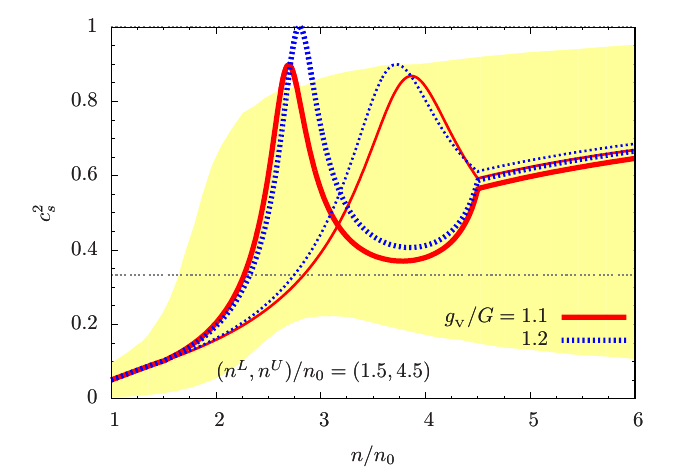}
   \includegraphics[width=6.0cm]{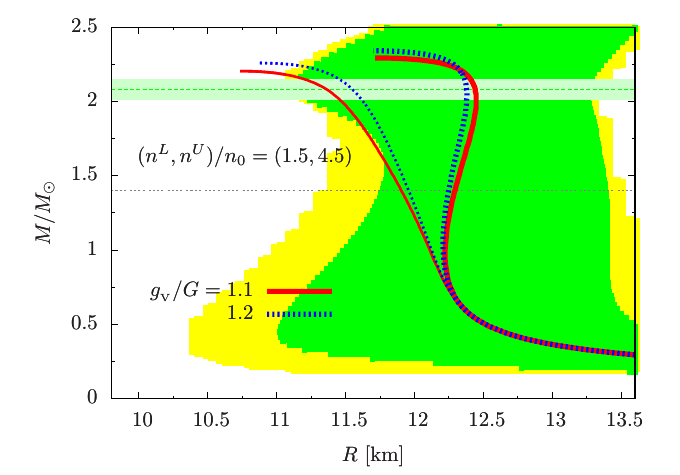}
   \includegraphics[width=6.0cm]{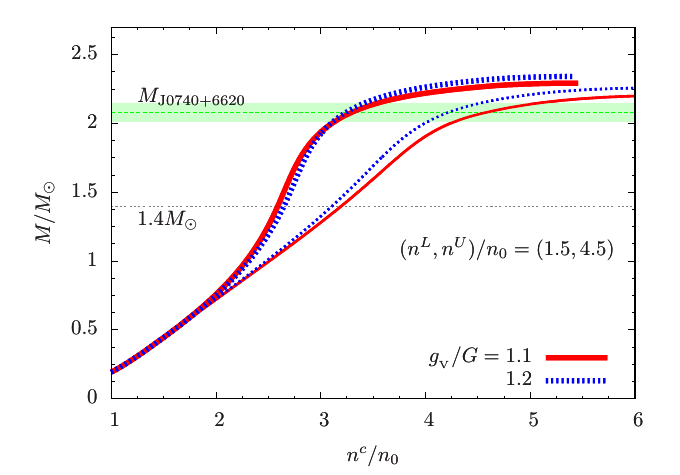} 
 \end{center}
 \end{minipage}
  \caption{  \footnotesize{
   \label{fig:M-R_vary}
    The relations among the sound velocity, mass-radius, and 
    \blueflag{central} density of neutron stars;
   $c_s^2$ vs. $n$ (upper panels), $M$ vs $R$ (middle panels), and $M$ vs. $n$ (lower panels), for matching densities $n^L$ =1.5 $n_0$ and $n^U$=3.5, 4.0, and 4.5 $n_0$.  
   For given $g_{_{\rm V}}=1.1$, and 1.2 $G$, the coupling $H$ has a lower (shown as bold) and an upper (shown as thin) bound (see Fig.~\ref{fig:gv-H-M-nL-nU}).  The plots for intermediate $H$ lie basically between the bold and thin lines.    
   The error bands are as those shown in Figs.~\ref{fig:M-R_A-D} and \ref{fig:cs2-nb_A-D}.	
}  }
\vspace{0.5cm}

\end{figure}

   Shown in Fig.~\ref{fig:M-R_vary} are $c_s^2$ vs. $n$ (upper panels), $M$ vs. $R$ (middle panels), and $M$ vs. $n$ (lower panels), for the matching densities $n^L=1.5 n_0$ and $n^U =(3.5, 4.0, 4.5) n_0$.   For given $g_{_{\rm V}}= 1.1$, and $1.2G$,
the lower (bold lines) and upper (thin lines) bounds on $H$ are as shown in Fig.~\ref{fig:gv-H-M-nL-nU}.    The plots for intermediate values of $H$ lie between these lines.

  There are clear correlations between the locations of the peaks in the sound velocity, the radii of $1.4M_\odot$ neutron stars, and the growth of the neutron star mass as a function of the central density.   
A peak in sound velocity at lower density indicates rapid stiffening which results in a larger radius.
In the upper panels of Fig.~\ref{fig:M-R_vary}, the locations of the peaks change by $\sim 1$-$1.5 n_0$ as $H$ is varied at given $g_{_{\rm V}}$, and the radii change by $\sim$ 0.3-0.7 km, as shown in the middle panels.
The central value of $H$ leads to $R_{1.4} \simeq$ 12.2-12.4 km, and with the variation of $H$,  the $M$-$R$ relations remain largely consistent with the NICER data.

\clearpage 

\bibliographystyle{aasjournal}
\bibliography{ref}

\end{document}